\documentclass[aps,preprint]{revtex4}%
\usepackage{amsfonts}
\usepackage{amsmath}
\usepackage{amssymb}
\usepackage{graphicx}%
\providecommand{\U}[1]{\protect\rule{.1in}{.1in}}

\begin{document}
\title{Comment on "Photon-assisted electron transport in graphene: Scattering theory analysis"}
\author{M. Ahsan Zeb}
\email{maz24@cam.ac.uk}
\affiliation{Department of Earth Sciences, University of Cambridge, UK.}

\begin{abstract}
It is argued that Trauzettel \textit{et al}. [Phys. Rev. B 75, 035305 (2007)]
made some mistakes in their calculations regarding the photon-assisted
transport in graphene that lead to \emph{uncoupled sidebands} and emergence of
step-like features in dG/dV (G is differential conductance and V is the bias
voltage). We discuss the relevant corrections and explain in detail how the
correct results are expected to be quite different than the incorrect ones.

\end{abstract}
\maketitle

Despite its simplicity the Tien-Gordon formalism of photon assisted
transport\cite{c1} does not usually allows exact analytical solution even in
the simplest case of a step-like ac potential profile. This is because the
inelastic scattering to the so called sidebands results in a set of equations
that are coupled in the sideband index. There are in principle infinite number
of sidebands so the system of equations cannot be solved exactly. This is true
regardless of the nature of the quasiparticles in a system, i.e., whether they
obey Schrodinger-like\cite{c2}, Dirac-like\cite{c3} or any other wave
equation\cite{c4} does not alter this situation. Although, the graphene system
shows many unusual properties, owing to the fact mentioned above it is still
very surprising to see an exact analytical solution of a photon-assisted
problem in graphene presented in ref\cite{c5}. Authors consider a region of
pristine monolayer graphene subjected to an ac signal in addition to a dc
voltage that shifts the bands relative to the rest of the system. They call
the two regions Gr(ac) and Gr(in) and consider transmission of electrons from
the former to the latter. They calculate analytical expressions for the
transmission amplitudes of various sidebands, present expressions for the
transmitted current through the interface of the two regions and differential
conductance G, and determine dG/dV as a function of applied bias voltage V in
the limit of zero temperature. A careful look at their calculations reveals
that they made a number of mistakes.
In the geometry they consider, in the
region Gr(ac), there will be a self consistent dynamic equilibrium
distribution due to the photon-assisted inelastic transitions. 
The authors
states they consider the incident wavefunction comprising of components at
energies of all the sideband relative to $\varepsilon$ weighted by the Bessel functions, which apparently seems correct. 
But, it can be easily seen that none of these "components" satisfies the wave equation. Nevertheless, since we can consider the incident particles at a single energy, this issue can be resolved simply by restating the problem. 
However, the following mistake makes all their
calculations incorrect so that their results become useless. The reflected and
transmitted wavefunctions need to account for the possible transitions to
lower and higher energies on emission or absorption of modulation
quantum/quanta of the ac signal. The reflected wavefunction $\Psi_{r}%
^{(ac)}(\overrightarrow{x},t)$ they consider is%
\begin{align*}
\Psi_{r}^{(ac)}(\overrightarrow{x},t)  &  =%
{\displaystyle\sum\limits_{m=-\infty}^{\infty}}
r_{m}J_{m}(\frac{eV_{ac}}{\hbar\omega})\Psi_{0,-}^{(ac)}e^{-i(\varepsilon
+\hbar m\omega)t/\hbar}\\
&  =\Psi_{0,-}^{(ac)}e^{-i\varepsilon t/\hbar}%
{\displaystyle\sum\limits_{m=-\infty}^{\infty}}
r_{m}J_{m}(\frac{eV_{ac}}{\hbar\omega})e^{-i\hbar m\omega t}%
\end{align*}
which even does not satisfy the wave equation unless the factor $r_{m}%
J_{m}(\frac{eV_{ac}}{\hbar\omega})$ equals Bessel function of order $m$ with
argument $(\frac{eV_{ac}}{\hbar\omega})$, i.e., if $r_{m}=1$ for all $m$ ( in
such a case it will represent particles only at energy $\varepsilon)$. The
correct form of the reflected wavefunction would be a linear combination of
components at all sideband energies with a phase factor $e^{-i(\frac{eV_{ac}%
}{\hbar\omega})\sin(\omega t)}$ ($=%
{\displaystyle\sum\limits_{m=-\infty}^{\infty}}
J_{m}(\frac{eV_{ac}}{\hbar\omega})e^{-im\omega t}$) due to the ac signal,
i.e.,%
\[
\Psi_{r}^{(ac)}(\overrightarrow{x},t)=%
{\displaystyle\sum\limits_{n=-\infty}^{\infty}}
r_{n}\Psi_{-,n}^{(ac)}e^{-i(\varepsilon+\hbar n\omega)t/\hbar}\times%
{\displaystyle\sum\limits_{m=-\infty}^{\infty}}
J_{m}(\frac{eV_{ac}}{\hbar\omega})e^{-im\omega t}%
\]
where $\Psi_{-,n}^{(ac)}$ is the solution of the wave equation without the ac
potential representing particles at energy $\varepsilon+\hbar n\omega$ moving
along negative x-direction and $r_{n}$ will give the amplitude for the
transitions to the $n$th sideband in reflected states. The transmitted
wavefunction $\Psi_{tr}^{(in)}(\overrightarrow{x},t)$ they considered
is\ given by $\Psi_{tr}^{(in)}(\overrightarrow{x},t)=$ $%
{\displaystyle\sum\limits_{m=-\infty}^{\infty}}
t_{m}J_{m}(\frac{eV_{ac}}{\hbar\omega})\Psi_{+,m}^{(in)}e^{-i(\varepsilon
+\hbar m\omega)t/\hbar}$, which is correct but contains unnecessary Bessel
functions $J_{m}(\frac{eV_{ac}}{\hbar\omega})$ that also means that the
amplitude for the transmission in $m$th sideband would be $t_{m}J_{m}%
(\frac{eV_{ac}}{\hbar\omega})$ instead of $t_{m}$ as considered by the
authors. Correcting for all above mistakes and omitting the Bessel functions
in transmitted wavefunction, boundary conditions lead to a set of coupled
equations that cannot be decoupled so analytical solutions for $t_{m}$ cannot
be found. Further, the relation
\[%
{\displaystyle\sum\limits_{n}}
\left\vert r_{n}\right\vert ^{2}+%
{\displaystyle\sum\limits_{n}}
\left\vert t_{n}\right\vert ^{2}=1
\]
, where the sums are only over the bands with the propagating modes, would
hold in this case instead of $\left\vert r_{n}\right\vert ^{2}+\left\vert
t_{n}\right\vert ^{2}=1$ for obvious reasons. In section-II of ref\cite{c5},
expressions for the current and conductance are given. Authors missed the
factors $\Psi_{+,m}^{\dag(in)}\sigma_{x}\Psi_{+,m^{\prime}}^{(in)}$ in the
summations in the expression for the current given in equation(15) that also affects the expression for the differential conductance given in equation(16). Finally, consider the step-like features in dG/dV presented
in figure(2). Authors attribute them to the vanishing density of states at the
Dirac point, which shows that these features may still persist in the correct
results. In the following we will explain how these steps arise in their
calculations and why they are not expected in the correct results.

The first point to note is the fact that all "sidebands" in their calculations
are independently contributing to the transmission, each like as if it were a
dc potential problem. There is \textit{no} photon-assisted transport at all.
At values of eV that are integer multiple of $\hbar\omega$, contribution of a
"sideband" is included to/excluded from G. This amounts to
\textit{"adding/removing"} particles at energy of that sideband (without
affecting other "sidebands"). Obviously, the current will increase sharply
whenever we would add more particles to it. The same is the origin of the step
like features in dG/dV. In a photon-assisted problem, where all sidebands are
strongly coupled to each other, the situation is very different. For example,
in photon-assisted problem transmission of particles through a new emerging
sideband reduces the sum of the total number of particles reflected and
transmitted at energies of other sidebands by the same amount. Further,
contribution of a sideband is usually much less than the central band, and
total transmission may even decrease if total reflection increases! So it
would be very unusual to have a \textit{sharp} increase in the number of
transmitted particles on emergence of a new contributing sideband (Note that
the sharp rise in current in the problem considered by Dayem and
Martin\cite{c6} has two different reasons. First, the absorption of energy
quantum/quanta from the microwave field, help electrons cross the energy gap
and make transition to empty conduction band from the filled valence band.
Second, the density of states is very large at the gap edges. ). Since the
correct value of transmission probability for any sideband is likely to have
strong and complex dependence on the energy and the propagation angle of
incident particles (Chiral Dirac fermions), the precise dependence of G or any
other quantity like dG/dV on V is hard to predict. However, due to the reasons
discussed above, one thing is obvious: the correct results are expected to be
significantly different than the incorrect ones presented in figure(2) in
ref\cite{c5}.

\end{document}